\documentstyle[11pt,paspconf]{article}
\input psfig.sty
\begin{document}

\title{Spectrophotometric Dating of Elliptical Galaxies in the Ultraviolet}
\author{Young-Wook Lee, Jong-Hak Woo, Sukyoung Yi}
\affil{Center for Space Astrophysics, Yonsei University, Seoul, South Korea}
\author{Jang-Hyun Park}
\affil{Korea Astronomy Observatory, Taejeon, South Korea}

\begin{abstract}
    The UV upturn phenomenon observed in elliptical galaxies is attractive 
for its potential value as an age indicator of old stellar systems. 
We present our most recent population models for the UV evolution 
of elliptical galaxies. We confirm that the dominant UV sources are either 
metal-poor or metal-rich hot horizontal-branch (HB) stars in local giant 
ellipticals, but we also note that the contribution from 
post-asymptotic-giant-branch (PAGB) stars overwhelms the UV 
spectrum at higher redshifts (look-back times). The model UV spectral 
energy distribution (SED) is therefore strongly affected by the current 
uncertainty of the mean mass of PAGB stars at higher redshifts. 
Fortunately, our models suggest that the far-UV observations at $z \ge 0.35$
could produce strong constraint on the PAGB mass, while observations at 
lower redshifts ($0.15 \le z \le 0.30$) would still provide constraints on the
models on the origin of the UV upturn.
Future observations of ellipticals from 
the {\it STIS/HST} and planned {\it GALEX} space UV facility will provide 
crucial database required for more concrete calibration of our UV 
dating techniques for old stellar systems.
\end{abstract}

\keywords{elliptical galaxies, UV upturn, hot HB stars}

\section{Introduction}
Recent observations on the origin of the UV upturn phenomenon of 
elliptical galaxies showed that hot HB and post-HB stars 
are the dominant UV sources in these systems (Ferguson et al. 1991; 
O'Connell et al. 1992; King et al. 1992; Bertola et al. 1995; Brown et al 1997). As demonstrated in our recent investigation (Yi et al. 1999), it is still
not clear, however, whether 
the dominant UV sources are metal poor (``metal-poor 
HB model"; Lee 1994; Park \& Lee 1997) or metal rich (``metal-rich HB model"; 
Bressan et al. 1994; Dorman et al 1995; Yi et al 1998). 
The ``metal-poor HB model" suggests that the dominant UV sources are very old,
hot metal-poor HB stars and their post-HB progeny,
although metal-rich PAGB stars also contribute some UV flux. 
In this picture, nearby
giant ellipticals are about 3 Gyr older than the Milky Way Galaxy, which would
imply non-zero cosmological constant from time scale test. On the other 
hand, the ``metal-rich HB model" suggests that the dominant UV sources are 
super metal-rich hot HB stars that experienced enhanced mass-loss and 
helium enrichment. If so, nearby giant ellipticals are similar in age 
to the Milky Way Galaxy, and there is no conflict with zero $\Lambda$ cosmology
from time scale test. 
Clearly, it is of considerable importance to understand the origin of the UV 
upturn. Considering this situation, we have recently investigated the evolution 
of UV flux with look-back time to see if two models predict substantially
different evolutionary patterns. We found that this would
indeed provide some strong observational test on the origin of the UV upturn 
phenomenon (Yi et al. 1999). 
The major uncertainty in these model calculations is the mass of 
PAGB stars, since it is only poorly constrained from direct observations.
In this paper, we report our progress in investigating the effect of PAGB mass in the
modeling of UV evolution of elliptical galaxies.

\begin{figure}[t]
%%\vspace{8cm}
\centerline{{\psfig{figure=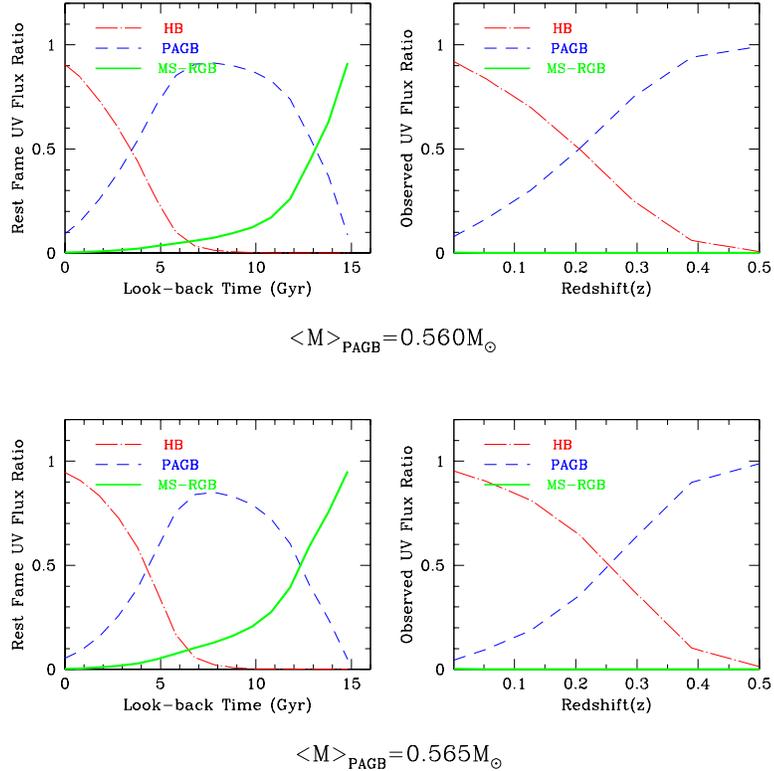,height=10cm}}}
\caption{Contribution of the various evolutionary phases to the total UV 
flux (1500\AA) of our models under ``metal-poor HB solution", plotted as a function of look-back time 
(redshift). Note the variation 
of the dominant UV source with look-back time. \label{fig-1}}
\end{figure}

\begin{figure}[t]
%%\vspace{8cm}
\centerline{{\psfig{figure=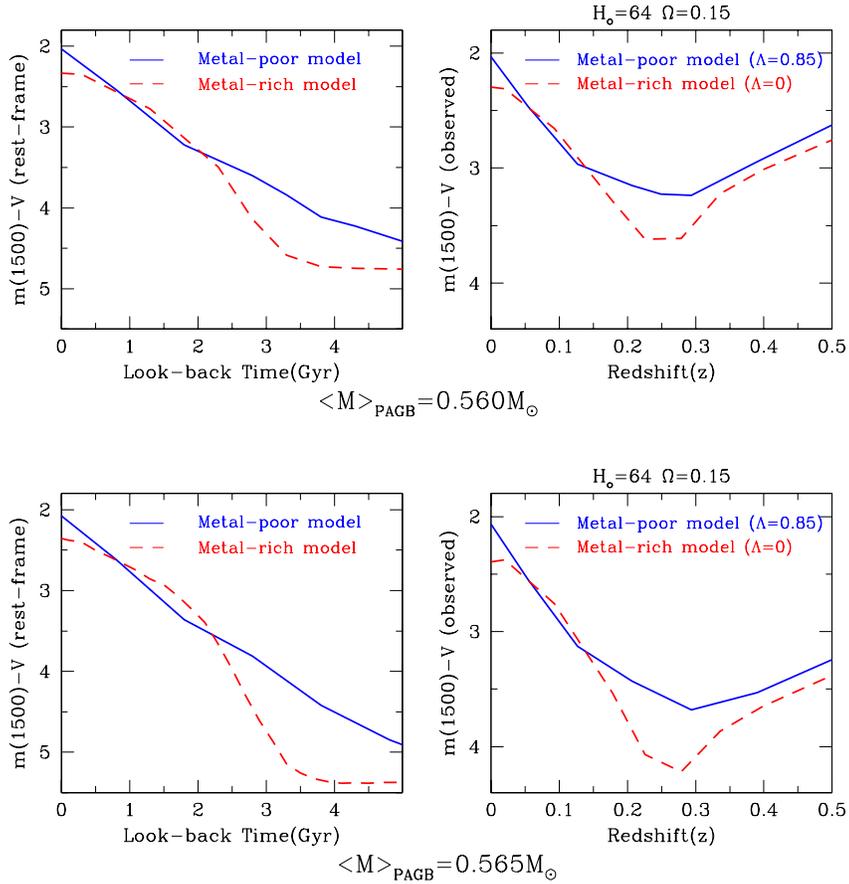,height=11cm}}}
\caption{The evolution of 1500\AA -V color index with look-back time \& 
redshift both for the ``metal-poor" and ``metal-rich" models. 
Input parameters in all models are calibrated 
so that they reproduce the IUE UV flux of nearby giant ellipticals. 
\label{fig-2}}
\end{figure}
\section{Variation of the Dominant UV source with Look-Back Tim}
    Our model calculations show that hot HB and post-HB stars contribute 
most of the UV light in nearby giant ellipticals (Fig. 1). As lookback 
time increses, however, PAGB stars are getting more dominat UV sources 
since the mean masses of helium burning stars are too high to be hot 
enough HB stars at younger ages. (At very large look-back times, 
hot main sequence stars eventually become the major UV source.)
    Since the contribution from PAGB stars overwhelms the UV spectrum 
at higher redshifts, the model SED is strongly affected by the current 
uncertainty of the PAGB mass as look-back time increases.

\section{1500\AA -V Color Evolution of Giant Elliptical Galaxies}
     We present, in Figure 2, the 1500\AA -V color evolution of 
giant ellipticals predicted by both
the ``metal-poor" and ``metal-rich" HB models under two assumptions 
regarding the PAGB mass. It is clear from Figure 2 that the 1500\AA -V 
is not strongly affected by the uncertainty of the PAGB mass 
for the nearby giant 
ellipticals since the PAGB contribution to the total UV flux is less 
than ~10\%. PAGB treatment becomes significant, however, for the 
ellipticals at $z \ge 0.15$. Fortunately, both ``metal-poor" and 
``metal-rich" models predict more or less 
the same 1500\AA -V colors at $z \ge 0.35$, 
and thus far-UV observations at these redshifts (cf. Brown et al. 1998) would
provide crucial constraint on the PAGB mass. Our models in Figure 2 suggest 
that the observations at lower redshifts ($0.15 \le z \le 0.30$) would still provide strong constraints on two models on the origin of the UV upturn.
Future observations of ellipticals from 
the {\it STIS/HST} and planned {\it GALEX} (Martin et al 1998) space UV 
facility will provide crucial database required for more concrete calibration 
of our UV dating techniques for old stellar systems.


\begin{references}
\reference Bertola, F., et al.\ 1995, \apj, 438, 680 
\reference Bressan, A., Chiosi, C., \& Fagotto, F. 1994, \apjsupp, 94, 63
\reference Brown, T. M., Ferguson, H. C., Davidsen, A. F., \& Dorman, B. 1997, \apj, 482, 685
\reference Brown, T. M., Ferguson, H. C., Deharveng, J.-M., \& Jedrzejewski, R. 1998, \apj, 508, L139
\reference Ferguson, H. C., et al.\ 1991, \apj, 382, L69
\reference King, I. R., et al.\ 1992, \apj, 397, L35
\reference Lee, Y.-W. 1994, \apj, 430, L113
\reference Martin, C., et al.\ 1998, \baas, 29, 1309
\reference O'Connell, R. W., et al.\ 1992. \apj, 395, L45
\reference Park, J.-H., \& Lee, Y.-W. 1997, \apj, 476, 28
\reference Yi, S., Demarque, P., \& Oemler, A. 1998, \apj, 492, 480
\reference Yi, S., Lee, Y.-W., Woo, J.-H., Park, J.-H., Demarque, P., \& Oemler, A. 1999, \apj, 513, 128
\end{references}
\end{document}